\documentclass[prb,twocolumn,showpacs,floats,eqsecnum,amsmath,amssymb]{revtex4}
\usepackage[dvips]{graphicx}
\def\be{\begin{equation}}
\def\ee{\end{equation}}

\def\bi{\begin{itemize}}
\def\ei{\end{itemize}}
\def\bn{\begin{enumerate}}
\def\en{\end{enumerate}}
\def\bea{\begin{eqnarray}}
\def\eea{\end{eqnarray}}
\def\no{\nonumber}
\def\ba{\begin{array}}
\def\ea{\end{array}}
\def\bd{\begin{displaymath}}
\def\ed{\end{displaymath}}
\def\la{\langle}
\def\ra{\rangle}

\begin{document}

\title{Enhanced superconducting proximity effect in
clean ferromagnetic domain structures}

\author{M. A. Maleki and M. Zareyan}

\affiliation{ Institute for Advanced Studies in Basic Sciences,
45195-1159 Zanjan, Iran}
\date{\today}

\begin{abstract}
We investigate the superconducting proximity effect in a clean
magnetic structure consisting of two ferromagnetic layered
domains with antiparallel magnetizations in contact with a
superconductor. Within the quasiclassical Green's function
approach we find that the penetration of the superconducting
correlations into the magnetic domains can be enhanced as
compared to the corresponding single domain structure. This
enhancement depends on an effective exchange field which is
determined by the thicknesses and the exchange fields of the
two domains. The pair amplitude function oscillates spatially
inside each domain with a period inversely proportional to the
local exchange field. While the oscillations have a decreasing
amplitude with distance inside the domain which is attached to
the superconductor, they are enhancing in the other domain and
can reach the corresponding normal metal value for a zero
effective exchange field. We also find that the corresponding
oscillations in the Fermi level proximity density of states as
a function of the second domain's thickness has an growing
amplitude over a range which depends on the effective exchange
field. Our findings can be explained as the result of
cancellation of the exchange fields induced phases gained by
an electron inside the two domains with antiparallel
magnetizations.
\end{abstract}


\maketitle

\section{Introduction}
In recent years mesoscopic ferromagnet-superconductor (FS) hybrid
structures have been studied very extensively.\cite{buzdin-rev}
These structures provide the possibility for the controlled
studies of the coexistence of the magnetic and superconducting
orderings in the presence of the phase coherent effects. The
motivation also has come from the interesting potential
applications, such as the proposal of the so called $ \pi $-SFS
Josephson junctions\cite{ryazanov} as the solid state
qubits.\cite{ioffe,mooij} The properties of a normal metal (N) in
close contact with an S, find superconducting characteristics due
to the penetration of the superconducting correlations (induced
order parameter) inside the normal metal. This proximity effect is
rather specific in an F metal, in which the interplay between the
spin splitting exchange interaction and the singlet
superconducting correlations makes the induced superconducting
order parameter to have a spatial damped oscillatory
behaviour.\cite{halterman} One of the manifestations of this
effect is that the density of states (DOS) of an F layer in
contact to S oscillates with varying the thickness and amplitude
of the exchange field of the F layer.\cite{zareyanprl,zareyanprb}
The oscillations of the proximity DOS in thin F layers with
different thicknesses have been observed in
experiments.\cite{kontos}
\par
The proximity effect in F is rather short range as compared to a normal metal
case in which the superconducting correlations can penetrate over the normal
coherence length $ \xi_N = v_{\text{F}} / T $. Here $ v_{\text{F}} $ is the
Fermi velocity and we use the units in which $ \hbar = k_B = 1 $. When the
exchange field $h$ is homogeneous, the induced order parameter contains the
singlet component and also a triplet component with the zero
projection $ S_z =0 $ of the total spin.\cite{bergeret1} These components
can survive in a clean F only over a short
distance $ \xi_F = v_{\text{F}}/h $ from the FS interface. In spite of this
observation there have been experiments\cite{petrashov1,giroud,petrashov2} in
which the superconducting proximity effect in F are reported which have an
amplitude and range much stronger and longer than what the above described
theory predicts.
\par
In a pioneering work Bergeret et
al.\cite{bergeret-rev,bergeret1,bergeret2,bergeret3} studied the possibility of
a long range proximity in ferromagnets with a local inhomogeneous
magnetization. They showed that in the presence of a nonhomogeneous spiral
magnetization vector close to the FS interface a triplet component of the
superconducting order parameter with projections $ S_z = \pm 1 $ can be
generated in addition to the singlet and the triplet $ S_z =0 $ components. These
triplet components are not affected by the exchange splitting and can spread
over a long distance of the normal coherence length $ \xi_N $. Very recently
new experimental evidences of the long range superconducting correlations in
ferromagnets have been reported. Sosnin et al.\cite{sosnin1} observed the
superconducting phase-periodic conductance oscillations of a ferromagnetic Ho
wire at a magnetic phase with a conical magnetization vector, in contact with
the Al superconducting ring. They found that the phase coherent oscillations
sustain for the Ho wire lengths up to 150nm at $T=0.27$K, which is much longer
than $ \xi_F \sim 6$nm. There has been also reports of observing  Josephson
supercurrent in SFS contacts with stronger CrO$_2$ ferromagnetic contacts at
even longer distances $0.3-1\mu$m, by group of Klapwijk in
Delft.\cite{klapwijk} In order to explain such long range effects, they refer
to the generation of the Bergeret's triplet component with $ S_z = \pm 1 $, due
to the inhomogeneity of the magnetization.
\par
In order to induce the long range triplet components $ S_z = \pm 1 $ in F from
a singlet superconducting condensate in S, the nonlocal noncollinear orientation
of the magnetization vectors is essential. The triplet correlations will not be
induced for a collinear configuration of the magnetization vector. In this paper
we study the possibility of a long range superconducting proximity effect in a
ferromagnetic structure with collinear magnetization vectors. We consider a
clean FS proximity system in which F consists of two layered
domains F$_1$ and F$_2$ whose magnetization vectors are pointed antiparallel
to each other and are separated by a sharp domain wall of negligible
width (see Fig. \ref{fig1}). Using the quasiclassical Eilenberger
equation,\cite{eilenberger} we calculate the superconducting pair amplitude
function  (order parameter) and the proximity DOS in the structure. The pair
amplitude function (PAF) spatially oscillates within each domain with a period
which is determined by the amplitude of the local exchange field.
\par
We find that while in F$_1$ the amplitude of the PAF oscillations decreases
with the distance from S, in F$_2$ it can increase over a distance which
depends on an effective exchange field determined by the thicknesses and the
exchange fields of F$_1$ and F$_2$. When F$_1$ and F$_2$ have the same
thicknesses and exchange fields, the PAF oscillations are amplified over whole
of F$_2$ and reaches to the value of the PAF for the corresponding NS
structure. We can understand this effect by noting that a quasiparticle
travelling through F with a given spin state will gain additional phases in
F$_1$ and F$_2$ due to the exchange field splittings. Since the exchange fields
are oriented antiparallel to each other the phases gained in
F$_1$ and F$_2$ will have opposite signs, leading to a phase cancellation and
suppression of the exchange fields effect. The similar effect was found before
in Josephson SFS junctions with F having a domain
structure.\cite{blanterhekking,bergeret5} We
further study the effect of this phase cancellation in the local proximity DOS
at the Fermi level. We show that it has an oscillatory behaviour with respect
to the exchange field and the thickness of F$_2$. The oscillations are
amplified up to the values of the thickness and exchange field which cancels
the phase effect of F$_1$ completely. At this point the DOS goes to zero
corresponding to the normal metal case. This kind of long range penetration
differs from the one introduced by Bergeret et al., as it has an oscillatory
behaviour and is superposition of the singlet and triplet components
with $ S_z = 0 $.
\par
The rest of this paper is organized as follows. In the next section we
introduce our model of the ferromagnetic domain structure (FDS) in
contact with an S and present solutions of the Eilenberger equation
for the quasiclassical Green's functions. In Sec.~\ref{sec3}, we
calculate the local subgap DOS and PAF. Sec.~\ref{sec4} is devoted to
the analysis of the PAF and the proximity DOS in terms of the involved
parameters. Finally in Sec.~\ref{sec5} we present the conclusion.

\section{The Model and basic equations\label{sec2}}
\begin{figure}
\vspace{0.2in}
\centerline{\includegraphics[width=7cm,angle=0]{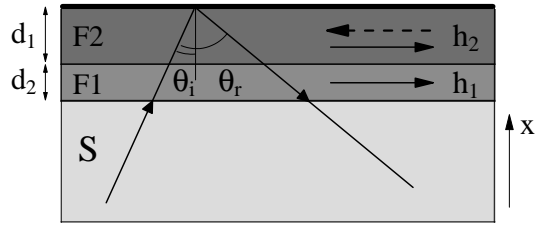}}
\caption{Schematic of the ferromagnetic domain structure in
contact with a superconductor. A classical electronic path is also
shown.} \label{fig1}
\end{figure}

The system we study is sketched in Fig. \ref{fig1}. An FDS consisting
of two layered domains F$_1$ and F$_2$ is connected to a
superconductor (S) in one side and is bounded on the other side by an
insulator or vacuum. F$_{1,2}$ have
thicknesses $d_1$ and $d_2$, respectively, with magnetization vectors which
are oriented antiparallel to each other. We characterize F$_1$ and F$_2$ by
mean field exchange splittings $h_1$ and $h_2$, respectively, which are
included in the Hamiltonian with different signs for antiparallel
orientation of the magnetization vectors. The thicknesses $d_{1,2}$ are
larger than the Fermi wave length $\lambda_{\text{F}}$ and smaller than the
elastic mean free path $\ell_{\text{imp}}$, which allows for a quasiclassical
description in the clean limit. We apply the collisionless Eilenberger
equation,\cite{eilenberger} which in the absence of the spin-flip
scatterings, is reduced to the following equation for each
of $ \sigma = \pm 1 $ spin directions:
\bea
&&\mathbf{v}_{\text{F}} \mathbf{ \nabla } \hat{G}_{\sigma}
(\omega_n ,\mathbf{v}_{\text{F}} ,\mathbf{r})
\label{b1} \\
&&+ \left[ \left( \omega_n - i \mathbf{\sigma} h (\mathbf{r})
\right) \hat{\tau}_3 + \hat{\Delta} ( \mathbf{r} )
, \hat{G}_{\sigma} (\omega_n ,\mathbf{v}_{\text{F}}
,\mathbf{r}) \right] =0. \no
\eea
The matrix Green's function for spin $\sigma$ has the form
\bea
\hat{G}_\sigma =\left(
\begin{array}{cc}
g_\sigma&f_\sigma \\
f_\sigma^\dag&-g_\sigma
\end{array}\right),
\label{b2}
\eea
where $g_\sigma$ and $f_\sigma$ are the normal and
anomalous Green's functions, respectively, which depend on
Matsubara's frequency $ \omega_n= (2 n+1) \pi T $
($T$ is the temperature), the direction of the Fermi velocity
$\mathbf{v}_{\text{F}}$, and the coordinate $\mathbf{r}$.
Here $ \hat{\Delta} (\mathbf{r}) = \Delta (\mathbf{r}) \hat{\tau}_1 $
is the superconducting pair potential matrix (taken to be real) and
$\hat{\tau}_i, (i=1,2,3)$ denote the Pauli matrices.
The matrix Green's function obey the normalization
condition $\hat{G}_\sigma^2 = \hat{1}$.
We assume that the exchange field is homogeneous in F$_{1,2}$ with
opposite signs, and vanishes inside S. Inside F$_{1,2}$ we
take $\Delta (\mathbf{r}) =0$. We also neglect the selfconsistent
variation of the pair potential close to the F$_1$S
interface, thus $\Delta (\mathbf{r})$ is constant inside S.
\par
In the ballistic limit we can solve Eq. (\ref{b1}) along a classical
electronic trajectory. A typical trajectory is shown in Fig. \ref{fig1}.
We parameterize the trajectory by a variable $ - \infty < \tau < \infty $.
An electron coming from bulk of the S, $ \tau = - \infty $, enters FDS at F$_1$S
interface, $ \tau =0 $, and subsequently F$_2$ at F$_1$F$_2$ interface,
$ \tau = l_{11} / v_{\text{F}} $. The electron undergoes a reflection at the
insulator and travels back along the returned segment of the trajectory toward
the S. In the returned path the variable $\tau$ at F$_1$F$_2$ and F$_1$S interfaces
is given by $ \tau = ( l_{11} + l_2 ) / v_{\text{F}} $ and
$ \tau = \tau_0 = ( l_1 + l_2 ) / v_{\text{F}} $, respectively. After passing
through F$_1$S interface, the electron returns to bulk of the S
where $ \tau = + \infty $. Here $l_{11}$ ($l_{12}$) is the length of the
trajectory segment inside F$_1$ at the coming (returned) part. The total length
of the trajectory inside F$_{1,2}$ are $l_1 =l_{11} +l_{12}$ and $l_2$,
respectively. To take into account the roughnesses at the insulator surface we
consider the reflections to be diffusive. Thus the directions of the coming and
returned parts of the trajectory are completely uncorrelated.
\par
The solutions of Eq. (\ref{b1}) are restricted by the boundary
conditions.These conditions determine the behaviour of the matrix Green's
function upon crossing different interfaces in the structure, and also its
limiting equilibrium values at bulk of the S. The matrix Green's function
approaches to the bulk value $\hat{G}_\sigma ({\text{bulk}})
= (\omega_n \hat{\tau}_3
+ \Delta \hat{\tau}_1) / \Omega_n$,\cite{kulik}
where $ \Omega_n = \sqrt{ \omega_n^2 + {\Delta}^2 }, $ at
the beginning and the end of a trajectory
($\tau=\pm \infty$), and $ \Delta(T) $ is the superconducting gap at the
temperature $T$.\cite{tinkham} For simplicity we assume that the
F$_1$S and F$_1$F$_2$ interfaces are ideally transparent. For F$_1$F$_2$
interface our assumption means that the corresponding domain wall is sharp
with a negligible width. In this case the boundary conditions are reduced
to the continuity of the matrix Green's function at these interfaces.
\par
By imposing the above mentioned boundary conditions on the solutions of
Eq. (\ref{b1}), we found that along a trajectory the normal Green's
function $g_\sigma$ inside FDS ($ 0 \leq \tau \leq \tau_0 $) does not
depend on $ \tau $. It depends only on the lengths $l_1$ and $l_2$, and
has the form
\be
g_{\sigma} (l_{11} , l_{12} ) = \tanh ( \beta_{n \sigma} + \alpha_n ),
\label{b3}
\ee
where
$ \beta_{n \sigma} =( \omega_{n \sigma}^{(1)} l_1
+ \omega_{n \sigma}^{(2)} l_2 ) / v_{\text{F}} $  and
$\alpha_n = \sinh^{-1} ( \omega_n / \Delta )$,
with $ \omega_{n \sigma}^{(j)} = \omega_n - i \sigma h_j $.
Inside S, the normal Green's function depends on $\tau$. For the
coming segment of the trajectory $ \tau \leq 0 $, we obtain
the following result
\bea
&&g_{\sigma} ( \omega_n ,\tau ,l_{11} , l_{12} )
\label{b4} \\
&&= \tanh \alpha_n + \frac{ \sinh ( \beta_{n \sigma} ) }{ \cosh \alpha_n
\cdot \cosh (
\beta_{n \sigma} + \alpha_n ) }
\cdot e^{2 \Omega_n \tau}. \no
\eea
For the returned segment of the trajectory, $ \tau \geq \tau_0 $, the
normal Green's function is related to the one for incoming segment
via $ g_{\sigma} ( \omega_n ,\tau_0 -\tau ,l_{11} , l_{12} ) $.
\par
For the anomalous Green's function we find the
following expressions in different segments of the trajectory
\begin{widetext}
\begin{eqnarray}
\begin{array}{l}
f_{\sigma} ( \omega_n ,\tau ,l_{11} , l_{12} ) =
\left\{
\begin{array}{l}
( 1- e^{2 \Omega_n \tau } ) / \cosh \alpha_n
+ \exp ( \beta_{n \sigma} + 2 \Omega_n \tau )
/
\cosh ( \beta_{n \sigma} + \alpha_n )
~~~~~~~~~~~~~~~~~~~~\tau \leq 0 \\ \\
\exp ( \beta_{n \sigma}
-2 \omega_{n \sigma}^{(1)} \tau )
/
\cosh ( \beta_{n \sigma}+ \alpha_n )
~~~~~~~~~~~~~~~~~~~~~~~~~~~~~~~~~~~~~~~~~~~ 0 \leq \tau \leq l_{11} / v_{\text{F}} \\ \\
\exp [ \beta_{n \sigma}
+ 2 ( \omega_{n \sigma}^{(2)} -
\omega_{n \sigma}^{(1)} ) l_{11} / v_{\text{F}} -2 \omega_{n \sigma}^{(2)} \tau ]
/
\cosh ( \beta_{n \sigma} + \alpha_n )
~~~~~~~ l_{11} /\tau \leq \tau \leq (l_{11} + l_2) / v_{\text{F}} \\ \\
\exp [ - \beta_{n \sigma} +2 \omega_{n \sigma}^{(1)} ( \tau_0 - \tau ) ]
/
\cosh ( \beta_{n \sigma}+ \alpha_n )
~~~~~~~~~~~~~~~~~~~~~~~~~~~~~ (l_{11} + l_2) /\tau \leq \tau \leq \tau_0 \\ \\
( 1- e^{2 \Omega_n ( \tau_0 - \tau ) } )
/
\cosh \alpha_n
+
\exp [ - \beta_{n \sigma} + 2 \Omega_n ( \tau_0 - \tau ) ]
/
\cosh ( \beta_{n \sigma} + \alpha_n )
~~~~ \tau \geq \tau_0.
\end{array}
\right.
\end{array}
\label{b5}
\end{eqnarray}
\end{widetext}
\noindent
\par
Eqs. (\ref{b3}-\ref{b5}) specify the matrix Green's function Eq. (\ref{b2}) in
all points of the structure. Using these results we can extract all the interested
physical quantities. In the next section we use these results to calculate
the local electronic DOS and the profile of the superconducting PAF.

\section{Proximity DOS and the superconducting pair amplitude function\label{sec3}}
The DOS is expressed in terms of the normal Green's function. To find
the DOS per trajectory, we have to calculate
\be
N (E , \tau , l_{11} , l_{12} )= \frac{N_0}{2}
\sum_{\sigma=\pm 1} \Re ~
g_\sigma ( \omega_n = -i E +0 ) \},
\label{c1}
\ee
where $N_0$ is the DOS at the Fermi level in the normal
state. In the following we will calculate the proximity DOS
for energies below the gap ($ \mid E \mid \leq \Delta$).
Replacing Eq. (\ref{b3}) into Eq. (\ref{c1}), for the DOS
inside the FDS ($ 0 \leq \tau \leq \tau_0 $) we find
\be
N ( E, l_{11} , l_{12} ) = \frac{N_0}{2}
\!\! \sum_{\sigma=\pm1} \sum_{n= -\infty} ^{+ \infty}
\!\! \pi \delta ( k_{ \sigma }^{(1)} l_1 + k_{ \sigma }^{(2)} l_2
- \Phi - n \pi ) ,
\label{c2}
\ee
where $ k_{ \sigma }^{(i)}= (E+ \sigma h_i)/v_{\text{F}} $ and
$ \Phi= \arccos (E/\Delta) $ is the Andreev phase. From
equation (\ref{c2}) we can see the fact that the subgap DOS is a sum
of $\delta$ functions resulting from Andreev bound states of electrons
of $E \geq 0$ (positive $n$'s) and holes of $E < 0$ (negative $n$'s). The
energies also follow from the quasiclassical quantization condition that
sets the argument of the $\delta$-function equal to zero. For a given spin
direction $\sigma$, the phase shift of the Andreev states caused by the
exchange field has two contributions $ \sigma h_1l_1/v_{\text{F}} $
and $ \sigma h_2l_2/v_{\text{F}} $, resulting from the phase gained by a
quasiparticle inside F$_1$ and F$_2$, respectively. For
antiparallel $ h_1 $ and $ h_2 $ these phases have opposite signs which
can lead to a cancellation of the exchange fields phase effects. As we
will see in the following this cancellation is responsible for a long
range penetration of the superconducting correlations inside the FDS.
\par
Inside S or $ \tau \leq 0 $ the subgap DOS is obtained by replacing
Eq. (\ref{b4}) in Eq. (\ref{c1}):
\bea
&&N ( E , \tau , l_{11} , l_{12} )= \frac{N_0}{2} \pi e^{ 2 \tau \sqrt{
\Delta^2 -E^2 } }
\label{c3} \\
&& \times \sum_{\sigma=\pm 1} \sum_{n= -\infty} ^{+ \infty}
\delta ( k_{ \sigma }^{(1)} l_1 + k_{ \sigma
}^{(2)} l_2 - \Phi - n \pi ), \no
\eea
where for returned part of the trajectory $ \tau \geq \tau_0 $,
we have the relation $ N ( E , \tau_0 - \tau , l_{11} , l_{12} ) $.
\par
The total DOS is obtained by averaging Eqs. (\ref{c2}-\ref{c3}) over all
different possible classical trajectories. This corresponds to an averaging
over Fermi velocity directions. To take into account the weak bulk
disorder, we include a factor $\exp (- v_{\text{F}} \tau_0 / \ell_{\text{imp}})$ in
our averaging.\cite{zareyanprb} Denoting the angles between the directions of the
coming and returned segments with the normal to the insulator
by $\theta_i$ and $ \theta_r $, respectively, the relations
$ l_{11} = d_1 / \mid \cos \theta_i \mid $, $ l_{12} = d_1 / \mid \cos \theta_r \mid $
hold. For diffusive reflections $ \theta_i $ and $ \theta_r $ are completely
uncorrelated, and we make the averaging by an independent integration over these
two angles. In this way from Eq. (\ref{c2}), inside FDS , we obtain
the following result for the total subgap DOS
\be
N (E)= \frac{N_0}{2} \sum_{\sigma=\pm 1}
\sum_{n= -\infty} ^{ + \infty}
P_{n  \sigma} ( E )
e^{ 2 i n \arccos (E/ \Delta) },
\label{c4}
\ee
where $ P_{n  \sigma} ( E ) = E_2^2 ( b + 2 i n
A_{ \sigma } ) / E_2^2 ( b ) $ with
$ b= b_1 + b_2 $, $ b_i= d_i / \ell_{\text{imp}}$ and
$ E_2 (z) = \int_{ 1 } ^{ \infty} d u \exp (- z u) / u^2 $
is the exponential integral of the second order. Here
\be
A_{ \sigma } = k_{ \sigma }^{(1)} d_1
+ k_{ \sigma }^{(2)} d_2 = k_{ \sigma } d,
\label{c5}
\ee
with $ k_{ \sigma }= (E+ \sigma h_{\text{eff}})/v_{\text{F}}$, is
an effective phase which is determined by the total thickness of
FDS $ d= d_1 + d_2 $ and an effective exchange field defined as
\be
h_{\text{eff}} = \frac{h_1 d_1 + h_2 d_2}{d_1 + d_2}.
\label{c6}
\ee
\par
From the result given by Eqs. (\ref{c4}-\ref{c6}) we conclude that the
proximity DOS of the FDS is equivalent to an effective single domain
ferromagnet with thickness $d$ and the exchange
field $ h_{\text{eff}} $. A similar result was obtain before for FDS
Josephson junctions.\cite{blanterhekking}
\par
By a same averaging over Eq. (\ref{c3}) the total subgap DOS is obtained
inside S
\be
N (E ,x)= \frac{N_0}{2} \sum_{\sigma=\pm 1}
\sum_{n= -\infty} ^{ + \infty}
Q_{n  \sigma} ( E ,x )
e^{ 2 i n \arccos (E/ \Delta) },
\label{c7}
\ee
where we have introduced the coordinate $x$ along the normal to the interfaces
with $ x<0 $ and $ 0 \leq x \leq d $ correspond to the points at the S and FDS
sides, respectively. Here $ Q_{n  \sigma} ( E ,x ) = E_2 ( b + 2 i n A_{ \sigma }
- 2 x \sqrt{ \Delta^2 -E^2 } / v_{\text{F}} )
E_2 ( b + 2 i n A_{ \sigma } ) / E_2^2 ( b ) $. While the subgap DOS at a given
energy $E$ vanishes in bulk of S, it can have a finite value close to F$_1$S
interface, due to the proximity effect. As we found above the proximity DOS
in FDS, Eq. (\ref{c4}), is constant.
\par
To calculate the PAF we have to do the same averaging over
Fermi velocity direction $\mathbf{v}_{\text{F}}$ as we did for the proximity
DOS, but this time over the anomalous Green's function given in
Eq. (\ref{b5}). In fact for PAF normalized in the way that goes to unity at
bulk of the S, we have the following relation:\cite{eilenberger}
\be
F (x,T)= \frac{ \lambda \pi T }{ \Delta } \sum_{\sigma=\pm1} \sum_{n}
\la f_{\sigma} (\omega_n ,\tau ,l_{11} ,l_{12} ) \ra,
\label{c8}
\ee
in which $\la \dots \ra$ denotes averaging over $\mathbf{v}_{\text{F}}$ and the summation
is taken over Matsubara frequencies. Here $\lambda$ is the electron-phonon interaction
constant.

\section{Results and discussions \label{sec4}}
Eqs. (\ref{c4}, \ref{c7}) and (\ref{c8}) express the proximity subgap DOS and
the superconducting order parameter in different regions of the FDS proximity
system. In the following we analyse the profile of these quantities as well
as their dependence on the effective exchange field Eq. (\ref{c6}). We pay
our attention to the weak ferromagnetic alloys like Pd$_{1-x}$Ni$_x $ with
$x$ of the order of $10 \%$, whose exchange field is estimated to be in the
range $ h= 5-20 $mev. For a Nb superconductor with
$ \Delta_0=\Delta(T=0) =1.4$mev we take $h_{1,2} / \Delta_0$ to be of
order $10$.\cite{kontos}
\par
\begin{figure}
\centerline{\includegraphics[width=9cm,angle=0]{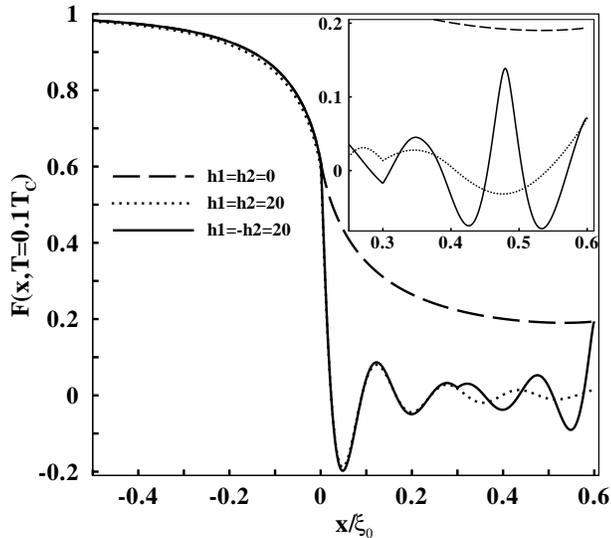}}
\caption{The pair amplitude function $F (x,T)$ normalized
to the bulk value, for the normal metal, single domain ferromagnet and
the ferromagnetic domain with $h_{\text{eff}} =0$ at $ T= 0.1 T_C $.
The thicknesses of domains are taken to be $ d_1 = d_2 = 0.3 \xi_0 $. The
exchange fields $h_{1,2}$ are written in the units
of $ \Delta_0 $. Inset shows the curves for ferromagnetic domain structure
with two nonzero effective exchange fields $ h_{\text{eff}} = -4 \Delta_0 $
(solid curve) and $ h_{\text{eff}} = 4 \Delta_0 $ (dotted curve).
In the solid curve, the maximum of PAF happens at $ x_p= 0.48 \xi_0 $.}
\label{fig2}
\end{figure}

Let us start with analysing the PAF. In Fig. \ref{fig2} the PAF
is plotted versus the coordinate $x$, for different values of the exchange
fields $h_{1,2}$ when the thicknesses $ d_1 = d_2 =0.3\xi_0 $ and
$ T= 0.1 T_C $. Here $ \xi_0 = v_{\text{F}} / \Delta_0 $ is the
superconducting coherence length in the clean limit.  At the S side close
to the F$_1$S interface the PAF is suppressed with an amplitude which is
almost independent of the values of $h_{1,2}$. Inside the FDS the PAF
depends strongly on the values of $h_{1,2}$ as well as on their relative
signs. While for a fully normal case of $h_{1}=h_{2}=0$ (dashed curve) the
PAF is exponentially decays with $x$ within the normal coherence
length $\xi_N = v_{\text{F}} / T $, it has a quite different behaviour for
nonvanishing $h_{1,2}$. For $ h_{1}=h_{2}=20\Delta_0 $ (dotted curve) the
FDS is transformed to a single domain ferromagnet, for which PAF oscillates
as a function of $x$ with a period $ \xi_F = v_{\text{F}} /h $ and a
damping amplitude. Here $ \xi_F \ll \xi_N $, since $ \Delta_0 \ll h $.
\par
An interesting new effect is appeared when the ferromagnetic part
is an FDS with the magnetization vectors oriented opposite to each
other. In Fig. \ref{fig2} the PAF is also plotted for the
case $ h_1 = -h_2 = 20 \Delta_0 $ (solid line), and in the inset for
the cases $ h_1 = 12 \Delta_0 $, $ h_2 = -20 \Delta_0 $ (solid curve)
and $ h_1 = 20 \Delta_0 $, $ h_2 = -12 \Delta_0 $ (dotted curve). In
these cases the PAF is an oscillatory function  in F$_1$ and F$_2$ with
a period which is given by the local exchange fields $h_1$ and $h_2$. The
PAF oscillations is always damping inside F$_1$. However the amplitude
of the oscillations can enhance in F$_2$ over a distance
from SF$_1$ interface, depending on the value
of $ h_{\text{eff}} $. We obtain that the distance at which the PAF
reaches its maximum amplitude is given
by $ x_p = ( 1 - h_{\text{eff}} / h_2 )d $. For the
case $ h_1 = -h_2 = 20 \Delta_0 $ the effective exchange
field $ h_{\text{eff}} = 0 $ and hence $ x_p = d $, which implies the
enhancement of the oscillations amplitude over whole the thickness
of F$_2$. In this case the phase cancellation effect is complete and
the PAF reaches the normal metal value at $ x=d $.
\par
For the cases shown in the inset of Fig. \ref{fig2}, the phase cancellation
occurs only partially. For $ h_1 = 12 \Delta_0 $, $ h_2 = -20 \Delta_0 $, the
effective field $ h_{\text{eff}} = -4 \Delta_0 $ is negative
and $ x_p =4/5 d < d $. In this case the amplitude of the PAF
in F$_2$ is enhancing for $ x < x_p $ and decreasing
for $ x > x_p $. For $ h_1 = 12 \Delta_0 $, $ h_2 = -20 \Delta_0 $, the
effective field $ h_{\text{eff}} = 4 \Delta_0 $ is positive
and $ x_p =6/5 d > d $. For this case the PAF is enhancing in whole
of F$_2$. Both the curves, in the inset, are reaching to the same value
at $ x = d $, which is always smaller than the normal metal value.
\par
We can understand the above mentioned results using the following simple picture.
We attribute the proximity effect in FDS to the presence of a macroscopic
number of the Cooper pairs which are escaped from the superconducting condensate
of the attached S and enter into the ferromagnetic region. For an homogeneous
exchange field in F region the Cooper pairs of electrons with opposite spins
gain an additional phase by the exchange splitting induced momentum. The exchange
field also tends to break the Cooper pairs by aligning the individual spins of
the electrons. These effects, respectively, lead to oscillation and suppression
of the Copper pair wave function, or correspondingly the PAF. In the case of the
FDS with antiparallel oriented magnetizations in F$_1$ and F$_2$, there is
another alternative possibility. Two electrons of a Cooper pair with opposite
spins can be in different domains. A fraction of these Cooper pairs with
electrons whose spins are parallel to the local exchange fields $ h_{1,2} $, can
survive from the above mentioned pair breaking. Such phenomena can lead to an
enhancement of the Cooper pair wave function, compared to an homogeneous F.
\begin{figure}
\centerline{\includegraphics[width=9cm,angle=0]{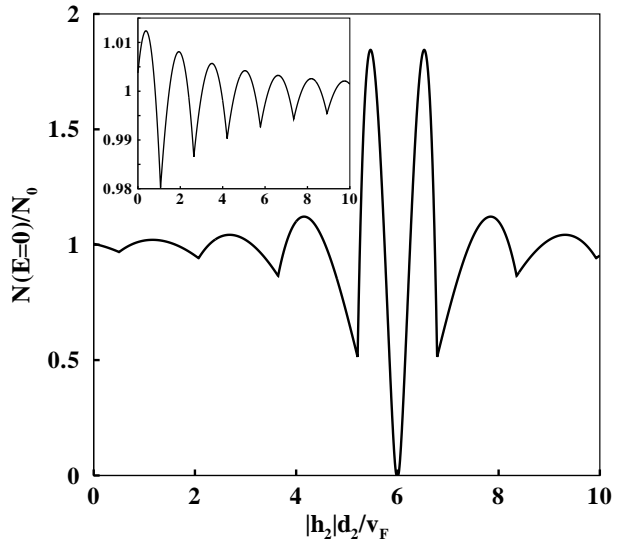}}
\caption{The behaviour of density of states inside the ferromagnetic domain
structure versus the dimensionless variable $h_2 d_2 /v_{\text{F}}$, for
$E=0$, $ d_1 = 0.3 \xi_0 $ and $h_1 =20 \Delta_0$. The exchange fields
in F$_1$ and F$_2$ are antiparallel. Inset: $h_1$ and $h_2$ are
parallel.}
\label{fig3}
\end{figure}
\begin{figure}
\centerline{\includegraphics[width=9cm,angle=0]{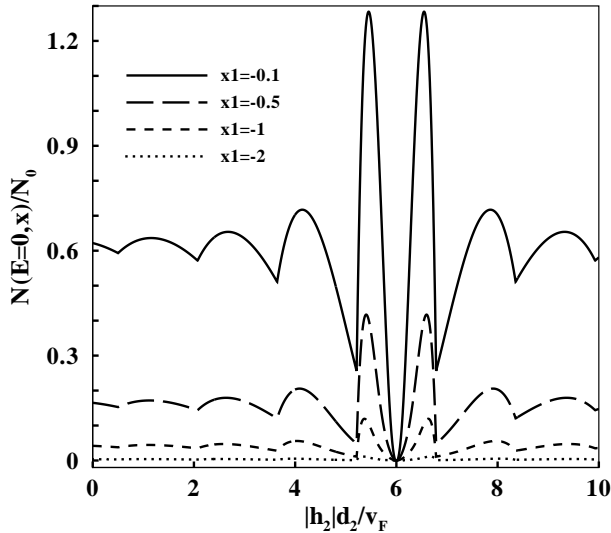}}
%\vspace{0.1cm}
\caption{The behaviour of density of states inside the superconductor
in terms of the dimensionless variable $h_2 d_2 /v_{\text{F}}$
for $E=0$, $ d_1 = 0.3 \xi_0 $ and $h_1 =20 \Delta_0$. $ x_1 = x / \xi_0 $ and
the exchange fields in F$_1$ and F$_2$ are antiparallel.}
\label{fig4}
\end{figure}
\par
This effect manifests itself also in the dependence of the proximity DOS of the
FDS on $h_2 d_2 /v_{\text{F}}$. In Fig. \ref{fig3} we plot the DOS at the Fermi
level ($ E=0 $) versus $h_2 d_2 /v_{\text{F}}$  for $ d_1= 0.3 \xi_0 $ and
$ h_1= 20 \Delta_0 $. The DOS oscillates around the normal state
value $N_0 $, with a period of $ \pi /2 $, similar to the DOS oscillations of
an homogeneous F.\cite{zareyanprl} But now in contrast to the homogeneous F
case for which the amplitude of the oscillations is damped
with $ h_2 d_2 /v_{\text{F}} $ (see the inset), the amplitude can grow over
a range of this quantity. In analogy with the PAF, this range is determined
by the effective exchange field $ h_{\text{eff}} $. In this case the DOS
amplitude enhances up to $h_2 d_2 /v_{\text{F}} =-6$, where it drops to zero
corresponding to the normal metal proximity DOS. At this
point $ h_{\text{eff}} = 0 $, which implies a full cancellation of the
exchange fields induced phases. We obtain the same dependence of the Fermi
level DOS on $ h_2 d_2 /v_{\text{F}} $, inside S. This is shown in
Fig. \ref{fig4} for different distances from F$_1$S
interface $ x $, when $ d_1= 0.3 \xi_0 $ and $ h_1= 20 \Delta_0 $. The DOS
oscillations with $ h_2 d_2 /v_{\text{F}} $ is the same as the DOS of
FDS, but now it is around a value which depends on $ x $. Increasing $ x $ from
zero (F$_1$S interface) this value decreases monotonically from $N_0$, and
approaches zero at the bulk of the S, $ x \gtrsim \xi_0 $.

\section{Conclusion \label{sec5}}
In conclusion we have studied the superconducting proximity effect
in a ferromagnetic domain structure which consists of two clean
layered domains F$_{1,2}$ separated by a sharp domain wall. Using
the quasiclassical Green's functions approach, we have calculated
the pair amplitude function (PAF) and the proximity density of
states (DOS). Our results show that in most cases the behaviour of
the FDS proximity structure is equivalent to a single domain
ferromagnet with thickness $ d=d_1+d_2 $ and an effective exchange
field $ h_{\text{eff}} = (h_1 d_1 + h_2 d_2)/(d_1 + d_2) $ . The
PAF inside the FDS has spatial oscillations with a period which is
inversely proportional to the magnitude of the exchange field in
each domain. We have found that while the oscillations in the
first domain is always damping with distance from the
superconductor, it is enhancing in the second one due to the
compensation of the spin splitting effects, by the two oppositely
oriented exchange fields. For $h_{\text{eff}} =0$, the
compensation is complete and the amplitude of the oscillations in
F$_2$ reaches to the value of the corresponding normal metal PAF.
\par
We further have found that the exchange splitting compensation has an
apparent effect in the oscillatory behaviour of the proximity DOS with
the thicknesses of the ferromagnets. Inside the FDS, the Fermi level
DOS oscillations around the normal state value has an enhancing
amplitude with the thickness of F$_2$ and vanishes at a thickness for
which $ h_{\text{eff}} = 0 $. Increasing further the thickness
of F$_2$, the DOS increases back and start to have a damped oscillations
around the normal state value. We have explained our results in a simple
picture based on surviving of a penetrated Cooper pairs inside
FDS, whose two electrons are in different domains with spin directions
parallel to the local exchange fields.

\section*{ Acknowledgments}
We thank Ya. M. Blanter and Yu. V. Nazarov for useful discussions.

\end{document}